# Green ammonia supply chain and associated market structure: an analysis based on transaction cost economics

Hanxin Zhao[*]

*Faculty of Technology, Policy and Management, Delft University of Technology, Jaffalaan 5, 2628BX Delft, the Netherlands*
[*] Corresponding author, *Email: h.zhao-1@tudelft.nl, Fax: +31 152783422*

# Abstract

Green ammonia is poised to be a key part in the hydrogen economy. This paper discusses green ammonia supply chains from a higher-level industry perspective with a focus on market structures. The architecture of upstream and downstream supply chains are explored. Potential ways to accelerate market emergence are discussed. Market structure is explored based on transaction cost economics and lessons from the oil and gas industry. Three market structure prototypes are developed for different phases. In the infancy, a highly vertically integrated structure is proposed to reduce risks and ensure capital recovery. A restructuring towards a disintegrated structure is necessary in the next stage to improve the efficiency. In the late stage, a competitive structure characterized by a separation between asset ownership and production activities and further development of short-term and spot markets is proposed towards a market-driven industry. Finally, a multi-linear regression model is developed to evaluate the developed structures using a case in the gas industry. Results indicate that high asset specificity and uncertainty and low frequency lead to a more disintegrated market structure, and vice versa, thus supporting the structures designed. We assume the findings and results contribute to developing green ammonia supply chains and the hydrogen economy.

# Key words

# 1. Introduction

## 1.1 Green ammonia in a hydrogen economy

The de-carbonization of fossil fuel-based energy systems is key to tackle climate change, to which, renewable energy development is a necessary step [1]. However, the intermittent renewable sources have challenged the adoption of renewable energy. Green hydrogen, produced from renewable energy, is expected to play a concerted effort to enable the energy transition [2]. With related technologies, such as power-to-gas and fuel cells, the transition to a hydrogen society offers an opportunity to decarbonize energy systems in all sectors [3]. Currently, due to the low density and high flammability, handling hydrogen faces greater challenges than challenges associated with conventional fuels [4]. Green ammonia, as a derivative of green hydrogen, has increasingly drawn attention as a clean energy source and a stable and economical means to storing and transporting hydrogen [5]. In addition, the conventional ammonia industry today accounts for 43% of global hydrogen consumption and contributes over 420 million tons of $CO_2$ annually. The transition to green ammonia production is seen as an opportunity to provide sufficient green hydrogen demand and achieve the industry de-carbonization [6, 7]. As a result, some countries and regions have moved a step. For example, Japan sees ammonia as the best means to import renewable energy and a roadmap has been developed to promote the demand to 50 million ton by 2050 [8]. About 10% of European ammonia will be produced using renewable energy by 2030 [9]. The Netherlands plans to start green ammonia import from 2024 [10]. Since the new industry is emerging, there is a need to study the future green ammonia supply chains and associate market development which are key to the creation of a green ammonia economy [11].

## 1.2 Research on green ammonia supply chains and the hydrogen economy

Recent studies have mainly contributed to techno-economic assessment of green ammonia production and associated supply chains. Some studies have evaluated technical performance of production process [12-14]. For example, Wen et al. proposed a multi-generation system with electricity, heat, and ammonia produced, and biomass-to-ammonia-to-power is used as an energy storage method [14]. Some studies have assessed environmental impacts of segments on the supply chain, such as production and transportation [15, 16]. For example, Bicer et al. assessed environmental impacts and energy efficiencies of ammonia produced by hydropower, nuclear, biomass, etc. using life-cycle assessment [16]. Some studies have concentrated on economic assessments of supply chains [17-21]. For example, Smith et al. evaluated the economic benefits of producing green ammonia from renewable local hydropower in Sierra Leone [17]. Guerra et al. analysed investments to production plants of green ammonia in Chile [19]. Zhao et al. evaluated impacts of hydrogen supply and economic incentives on green ammonia production and investment [20, 21]. In the literature search, we found studies on the holistic supply chains, especially on the market development at a macro level are limited.

Since green ammonia industry is a part of hydrogen energy transition, more studies have contributed to the hydrogen economy development. Some studies have focused on reviewing institutional conditions for the hydrogen energy transition [22, 23]. For example, Velazquez Abad et al. reviewed definitions, standards and policy initiatives relevant to green hydrogen [22]. Another group of studies have discussed the feasibility of building a hydrogen economy [24-26]. For example, Li et al. evaluated economic feasibility of applying green hydrogen in the road transport sector in China [26]. There are also studies that have evaluated policy effects for the adoption of green hydrogen [27, 28]. For example, by applying a computable general equilibrium model, Bae et al. found price subsidy would exert positive impacts on hydrogen demand in Korea [28]. Hydrogen market development is increasingly paid attention [29-33]. Particularly, the creation of hydrogen market by producing green hydrogen using excessive renewable power has been explored [29-31]. For example, Burg et al. concluded that it is not economically feasible to integrate hydrogen market into electricity market in Denmark, and the net present value is negative on both sides [31]. Some studies have also discussed potential markets especially in the mobility sector [32, 33]. These studies have mostly discussed hydrogen policies from aspects of regulation, subsidy, etc., and an increasing number of studies have started to explore the future hydrogen market which is crucial to the hydrogen supply chain and economy development. However, studies on hydrogen supply chains and associated markets from an industry perspective are still not sufficient. Particularly, governance structure of markets, or known as market structure by many studies [34], which is key to the industry development has not been paid much attention.

## 1.3 Research goal and contribution

In summary, green ammonia is expected to play a key role as a clean fuel and energy carrier in the hydrogen energy transition. The development of associated supply chains is crucial to the green ammonia industry creation and growth, since energy systems are essentially a supply chain comprising of segments from production/imports to end use [1]. However, recent studies on green ammonia from a higher-level industry perspective are limited. In particular, the industry performance are highly dependent on market structure (i.e. market governance structure), however, studies on market structure of green ammonia and even hydrogen are rare. This paper aims to study green ammonia supply chains with a focus on the development of associated market structures relating to vertical integration and disintegration of the industry. Considering modern supply chains are not only technical systems, but also involves social aspect (e.g. policies, markets, etc.), this study is carried out from the socio-technical perspective. The main contributions are summarized as follows:
(1) This study contributes to exploring the holistic supply chain development for establishing the green ammonia industry, which is distinct with previous studies on green ammonia and hydrogen. It explores the potential architecture of green ammonia supply chains, and concentrates on discussing market structures and associated market development. We consider the results are not only of significance to the green ammonia industry, but also shed light on the market development in a hydrogen economy.
(2) Market structures and contracts are analysed based on transaction cost economics and lessons from the oil and gas industry. Three market structure prototypes are developed for the infancy and later development stages. In addition, the rationality of the designed structures are estimated in this study by quantitative analysis with results from a multi-linear regression model, which is few in discussing market structures.

## 1.4 Structure of the paper

The remainder of this paper is arranged as follows: the framework for analysing socio-technical energy systems, and transaction cost theory are introduced in Section 2. Section 3 discusses key results of the supply chain development, including: technical infrastructure, market structures, actors and associated contracts, and ways to accelerate the market emergence. Section 4 presents the conclusions, policy implications and limitations of the work.

# 2. Methods

## 2.1 Framework for analysing socio-technical systems

Energy transitions are defined as long-term structural changes in energy systems [35]. In addition, energy system is essentially a supply chain, known as a socio-technical system, as it is not only a technical matter, values of individual actors, policies, regulations and markets also shape the system [1, 36]. Fig. 1 shows the framework applied in this study for analysing socio-technical systems based on previous work [37, 38]. The framework derives from the theory of complex system engineering which addresses not only challenges in technical dimension, but also multi-actor complexity of socio-technical systems [37]. The framework builds on three key pillars, including: technical system, actors and institutions. Technical system refers to technical components in energy systems. Institutions are the devised rules that shape human interactions, while actors are the entities making decisions and participating in the process. The connections emphasizes the interactions between each other. Actors build and operate technical systems, which in turn influence actors' decisions. Institutions such as norms, strategies influence actor behaviors, which in turn reshape institutions. The three pillars and interactions among them should be considered simultaneously when designing or analysing socio-technical systems. Therefore, these are considered in this study in analysing green ammonia supply chains which are socio-technical systems. In addition, transaction cost theory is applied for analysing market structures which are introduced in the next section.

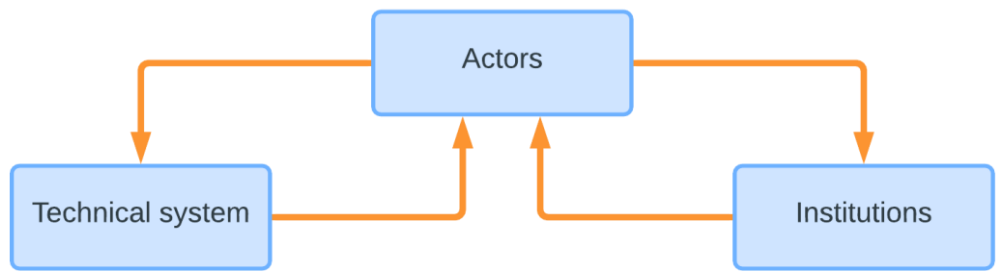

**Fig. 1.** Framework for analysing socio-technical systems [37]

## 2.2 Approach for analysing market structures

The concept of 'transaction costs' was first raised by Coase, and further developed as the theory of transaction cost economics (TCE) by Crawford et al., Riordan et al., Williamson, etc. [39-42]. According to TCE, conducting any transaction comes at a certain cost which may come from gathering information, bargaining, contracting, enforcement of contract agreements, etc., namely, transaction cost [41]. Other than considering markets as the best means for resource allocation, market contracts are assumed as ubiquitous incomplete due to the incapability of identifying all contingencies and vague descriptions on many key features, therefore does not apply to all types of transactions. In essence, TCE is an attempt to address organizational efficiency in economic transactions by interpreting vertical boundaries of firms and the needs for vertical integration.

A transaction should align with the governance structure which minimizes the transaction cost. The governance structure is classified into three types, including: market, hybrid, and hierarchy (or firm), as shown in Fig. 2, where $k$ represents contractual hazard, and $s$ represents safeguard [41]. Asset specificity, uncertainty and frequency are regarded as main transaction attributes that influence transaction costs which result in different governance structures. Definitions of the attributes are summarized in Table 1 [41].

Market is the ideal governance structure when there is no hazard driven by main transaction attributes. It principally refers to spot markets where transactions are coordinated by price mechanism [43]. With hazards enlarged, hybrid and hierarchy structures become more optimal when safeguards are elicited, otherwise it can lead to contractual breakdown (i.e. the status of unrelieved hazard in Fig. 2). If administrative safeguards as additional transaction cost turn out to be more economic, hierarchy becomes a preferable alternative. In this case, transactions are taken out of market and organized under unified ownership by using administrative commands. In other words, the business is vertically integrated within a firm instead of trading in a market. Vertical integration is considered by TCE as an efficient response to uncertainties and contractual incompleteness. Hybrid is an intermediate alternative between the market and hierarchy by means of additional contractual support from markets, such as joint ventures, partial ownership, contracts of varying duration and specificity, etc. [34].

**Table 1**
Definition of transaction attributes [41]

| Transaction attribute | Definition |
|---|---|
| Asset specificity | The degree of certain assets needed to support particular transactions. |
| Uncertainty | Disturbances to which transactions are subject. |
| Frequency | The degree that transactions recur. |

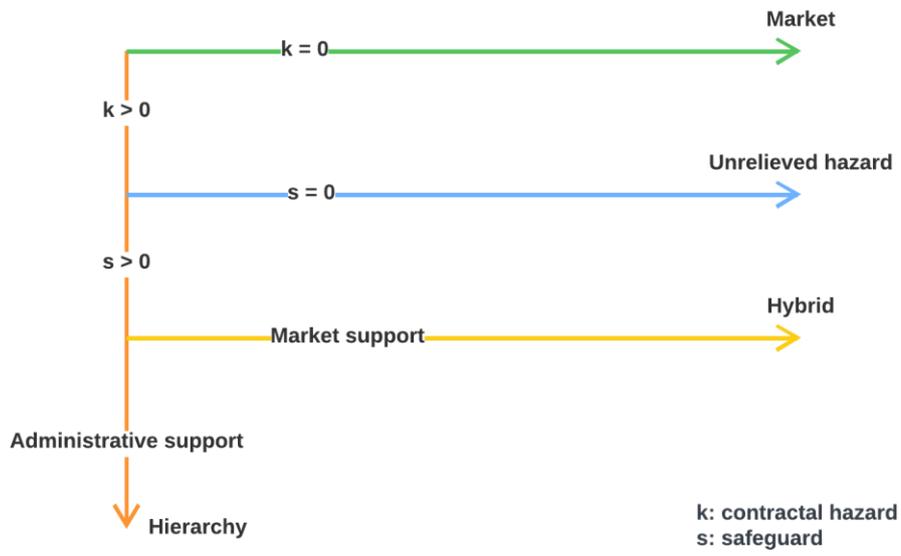

**Fig. 2.** Simple contractual schema [41]

# 3. Results and discussions

## 3.1 Supply chain architecture

Potential architecture of green ammonia supply chains is proposed in Fig. 3, after a detailed analysis of the infrastructure in the oil and gas industry and studies on hydrogen supply chains [44-46]. It comprises of upstream and downstream supply chains. The upstream supply chain concerns the supply side consisting of segments of hydrogen production, ammonia production, and ammonia transportation. Hydrogen production plants can be either installed within renewable power stations, or close to substations where renewable power is converged. Ammonia production is preferably set up close to hydrogen plants, or a little distant from hydrogen plants, for which hydrogen transportation should be considered. Subsequently, hydrogen is stored as ammonia for further large-volume and long-distance energy distribution. The downstream concerns the demand side, and there are potentially two types of supply chains. One concerns scenarios where ammonia can be directly consumed, such as being used as a fertilizer, intermediate chemical product, and clean fuel for power generation and heat supply [47]. The other category of downstream supply chain mainly serves for hydrogen applications, such as hydrogen for heating, mobility and industrial use. In this case, ammonia reform process is required to convert ammonia to hydrogen and import it to local hydrogen distribution networks, such as hydrogen pipeline grids or road transportation networks. Besides, It is also appropriate to define storage and transportation as the midstream supply chain similar to some practices that classify oil or gas supply chains [44]. In this case, supply chains are divided into three parts.

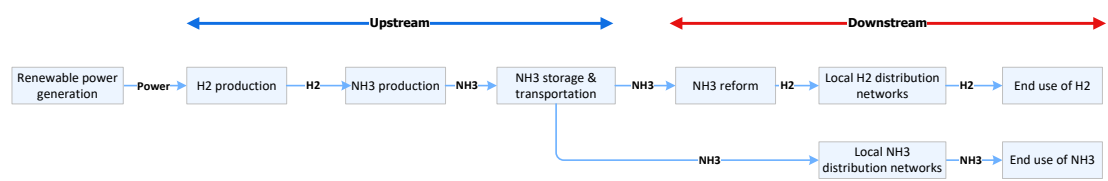

**Fig. 3.** Potential architecture of green ammonia supply chains

## 3.2 Market structure

Market structure describes relations among buyers and sellers in an industry, and is differentiated based on a set of factors contributing to the degree and nature of competition for products and services, such as availability of multiple players, vertical integration extent of the industry, etc. [34]. Given that supply chains are socio-technical systems, market structures are discussed by considering technical system presented in Section 3.1, as well as actors, institutions and related interactions. Potential market structures and associated contracts are developed based on TCE as well as lessons and practices mainly from the oil and gas industry [41, 44, 48-50]. Three types of market structure are developed for different development stages. Since the designed structure prototypes are not constant, but can vary with different conditions, potential mixed structures derived from the three prototypes are also discussed. Besides, it should be noted that, since renewable power generation is beyond the upstream supply chain, contracts or vertical integration between renewable power generation and hydrogen production is not involved. In addition, ways to accelerate market emergence are also analysed.

### 3.2.1 Integrated structure

Given considerable capital investments required and relatively limited and unstable demand, green ammonia industry in its infancy features high asset specificity, high uncertainty and low frequency. As a result, vertical integration is an efficient practice to address contractual hazards driven by transaction attributes, according to TCE. The practice was also adopted in the oil and gas industry in the early phase [48, 51]. However, the integration of both upstream and downstream supply chains should be avoided to prevent absolute monopoly. For example, the emergence of giant companies covering the business from oil exploration to petrol station operation has significant implications to the efficiency of oil markets [48]. This was eliminated on a large scale in the subsequent gas industry [51].

Fig. 4 shows a potential integrated market structure, where the upstream and downstream supply chains are both vertically integrated. It should be noted that it is also possible to integrate hydrogen production into renewable power generation in the early stage. Since as mentioned, we focus on discussing market structure on the basis of green ammonia supply chains, this part is not presented. Apart from solely investing in the upstream or downstream supply chain, hybrid governance can also be applied, according to TCE. For example, a joint investment from multiple large energy enterprises is proposed to share risks among participants, which is adopted in the hydrogen supply chain development in the Netherlands [10]. This can be done by signing a Joint Venture Agreement (JVA), by which, stakeholders can jointly invest and operate the integrated supply chain to be able to optimize the entire upstream supply chain. State-owned enterprises (SOEs) or large energy companies can take more initiatives in the early stage on both upstream and downstream supply chains to improve credibility. In addition, long-term contracting (e.g. between ammonia producers and transportation operators) is also a valid practice to share risks between buyers and sellers, which is further discussed in the following paragraph.

Similar to the practices in such as oil, gas and renewable power industries [48, 49, 52], the creation of bilateral contracts between producers and consumers is proposed to help with the emergence of a green ammonia market. This can be done by signing long-term Sale and Purchase Agreement (SPAs). A SPA should be required to run for a long term (e.g. 15-20 years) and include take-or-pay (TOP) provisions which obligate the minimum volume that buyers should take with a relatively fixed price to share risks between both sides and guarantee investment recovery. In addition, government participation is essential by setting regulatory and financial institutions to support market creation and development of supply chains, e.g. granting licenses for hydrogen and ammonia production and ammonia reform, approving SPAs, providing subsidies, etc. which refer to approaches in the gas industry [53]. Besides, main contracts formulated are summarized as follows.

(1) Joint Venture Agreement (JVA) is the contract between multiple stakeholders. It can be carried out either through written contracts or through setting up a separate corporation entity. Main elements include: types of joint venture, duties, obligations, etc.
(2) Sale and Purchase Agreement (SPA) is the long-term contract between buyers and sellers. Main elements in SPAs include: term, TOP clauses with minimum commitment described, price, transport arrangements (if it is used for ammonia trades), etc.

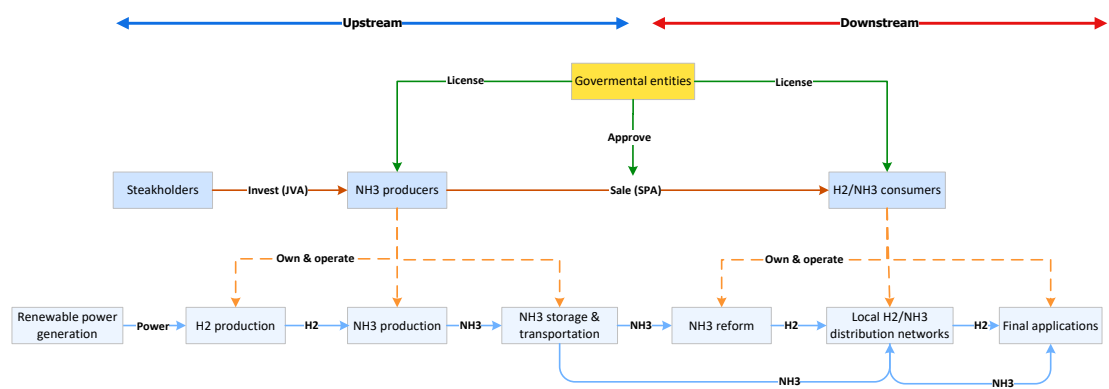

**Fig. 4.** A potential integrated market structure

## 3.2.2 Disintegrated structure

With the progress of business and technology, uncertainty will mostly decrease and frequency of transactions will increase in next stages. In addition, since the bilateral contracts limit the ability of trading products or providing services to other customers, the surplus can grow and finally becomes a serious issue due to the mismatch between supply and demand with the industrial scale increased. As a result, asset specificity will be reduced. According to TCE, market governance mode will be increasingly needed. This can lead to the emergence of liberal markets and the transition to a disintegrated market structure characterized by a vertical disintegration between segments on the supply chain.

From the perspective of enterprises, buyers and sellers can renegotiate the bilateral contracts when hitting the end to add more flexibility (e.g. a decrease in the term and TOP volume), and even organize short-term trades bilaterally. More importantly, government intervention is indispensible to implement the supply chain restructuring by introducing competition mechanisms. We propose practices by referring to the restructuring in the oil and gas industry [44, 50]. Regulatory intervention is needed to force companies to reduce their business scale and open third-party access to some segments necessary to break down. In addition, it is necessary to cultivate and develop short-term and spot markets, not limited to ammonia trades, but also for other key segments, such as transportation, ammonia reform, etc. Specifically, government need to take responsibility of, such as establishing trading hubs, enacting related regulations, overseeing contracts, granting licenses (Fig. 5 describes licenses for key segments, including: licenses for operating hydrogen production, ammonia production and reform, as indicated by the green arrows), etc. These practices aims to realize efficient allocation and break up monopoly. It is particularly necessary to avoid a further integration which can be observed in the oil industry [48]. This can occur in the case that some participants may not see cost-cutting as a top priority (e.g. some SOEs), or seek for extending their influences by integrating additional segments to cover the loss due to the inefficiency of existing businesses.

Fig. 5 shows main features of a disintegrated structure. It is characterized by a separate investment and operation of most of segments by individual actors. In the upstream supply chain, hydrogen production can detach from ammonia production. Hydrogen is traded between hydrogen producers and ammonia producers via Hydrogen Feedstock Agreements (HFAs). Ammonia transport can also be restructured to be able to serve for multiple ammonia buyers or sellers via an Ammonia Transportation Agreement (ATA). In the downstream, local hydrogen operators is only responsible for ammonia reform, and resale hydrogen to local hydrogen consumers via Hydrogen Sales Agreements (HSAs). Local hydrogen distribution can also be operated separately serving for buyers or sellers via Hydrogen Distribution Agreements (HDAs). It should be noted that the restructuring will create multiple subsidiary markets on the supply chain, e.g. ammonia transportation trading markets, due to the access to the third parties. Details of these subsidiary markets are not highlighted in Fig. 5. Besides, main contracts formulated are summarized as follows.

(1) Ammonia Transportation Agreement (ATA) is the contract with ammonia transport operators for long-distance ammonia delivery and storage (if needed). Main elements include: transportation fees, storage fees, minimum commitments, reservation fees, etc.
(2) Hydrogen Distribution Agreement (HDA) is the contract with hydrogen distribution operators for local hydrogen distribution. Main elements include: distribution fees, minimum commitments, reservation fees, etc.
(3) Hydrogen Feedstock Agreement (HFA) is the contract between hydrogen buyers and sellers for green hydrogen supply. Main elements include: price, payment terms, TOP clauses (if needed), pass-through of SPA liabilities (if needed), etc.
(4) Hydrogen Sales Agreement (HSA) is the contract between local hydrogen operators and consumers. Main elements include: price, commitments of buyers (TOP clauses if needed), payment term, etc.

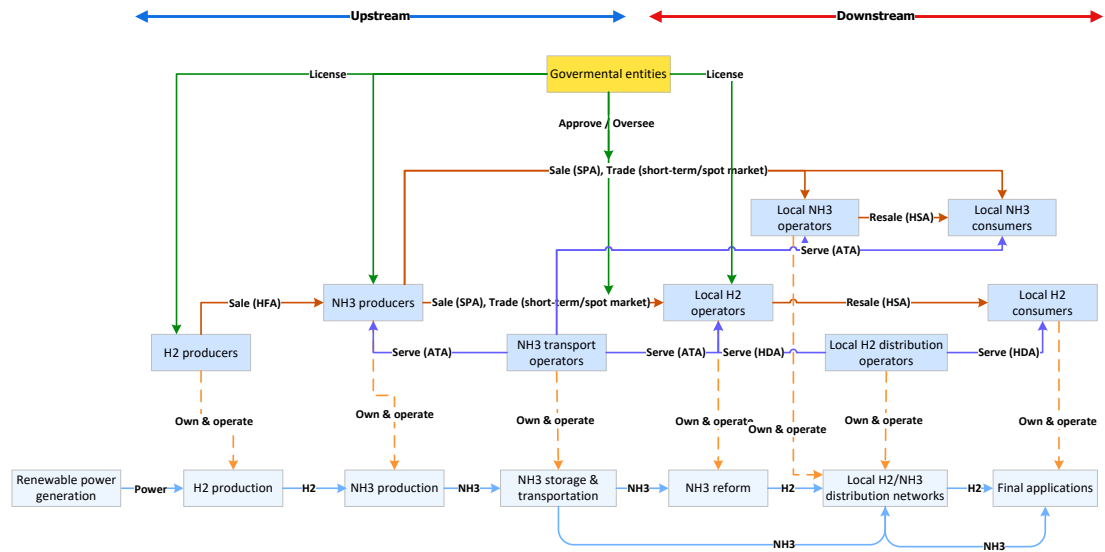

**Fig. 5.** A potential disintegrated market structure

### 3.2.3 Competitive structure

As mentioned in the above section, vertical disintegration on the supply chain is inevitable and necessary with asset specificity and uncertainty decreased, and transaction frequency increased. The competitive structure refers to a further vertical disintegration in main segments on the supply chain. It features a separation between asset ownership and production activities, such as hydrogen production and ammonia reform. This can be observed in the gas industry in the late stage [49]. Specifically, producers tend to outsource production activities to third-parties who own the facility and provide production service.

Fig. 6 shows a potential competitive structure. Hydrogen production can be invested and operated by hydrogen production operators who provide electrolysis services to hydrogen producers via Electrolysis Agreements (EAs). In this case, hydrogen producers pay tolling fees to electrolysis operators but owns the asset of hydrogen, and electrolysis operators can serve for multiple customers to improve facility utilization and increase revenues. Similarly, ammonia synthesis can be separated as a tolling service for ammonia producers via Ammonia Synthesis Agreements (ASAs). Meanwhile, aggregators can also emerge. Aggregators can collect hydrogen or ammonia from small-scale producers who might face difficulty in managing a trade (e.g. incapable of using transportation services) and resell it to buyers. In the downstream, ammonia reform can be operated separately by operators which provide reform services for local hydrogen operators via Ammonia Reform Agreements (ARAs). In addition, ammonia storage and distribution can be operated by specific actors providing services for local ammonia operators via Ammonia Logistics Agreements (ALAs). In addition, short-term and spot markets can be further developed, with fewer buyers and sellers willing to stay in long-term contracts. Besides, main contracts are summarized as follows.

(1) Ammonia Logistics Agreement (ALA) is the contract with the ammonia logistics

operators for local ammonia storage and transport. Main elements include: storage and distribution fees, reservation fees, etc.

(2) Ammonia Reform Agreement (ARA) is the contract between ammonia reform plant users and facility owners. Main elements include: service fees, minimum commitments, reservation fees, etc.

(3) Ammonia Synthesis Agreement (ASA) is the contract between ammonia production plant users and facility owners. Main elements include: service fees, minimum commitments, reservation fees, etc.

(4) Electrolysis Agreement (EA) is the contract between electrolysis facility users and owners for hydrogen production. Main elements include: service fees, minimum commitments, reservation fees, etc.

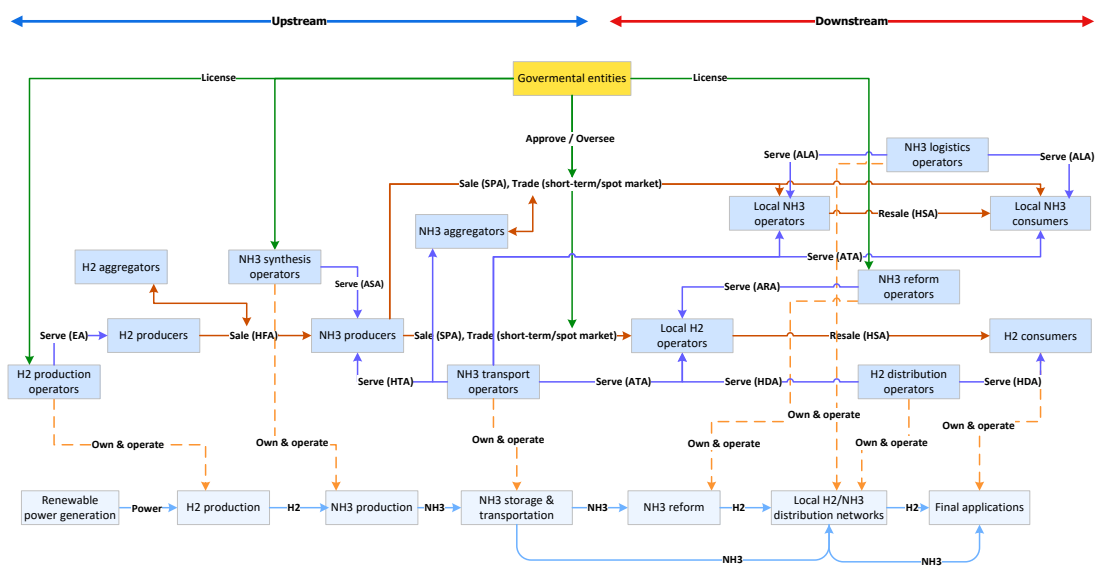

**Fig. 6.** A potential competitive market structure

## 3.2.4 Mixed structure

The three market structures discussed above are prototypes corresponding to main features of different development phases. Therefore, the market structure can vary with actual conditions (e.g. political, economic and geographical factors, etc.). For example, Fig. 7 shows a market structure where ammonia production is integrated with storage and transportation, but separated from hydrogen production. In the downstream, ammonia reform is still integrated with distribution segment, but final applications are operated independently. In addition, short-term markets are possible to emerge before shifting to the disintegrated structure. This can be seen as a mixed structure combining features of integrated and disintegrated structures. A potential combined disintegrated and competitive structure could be the case where key segments (e.g. hydrogen production and ammonia reform) can be partially operated by the facility owners to produce and sale their own share of products, and the rest of capacity can be traded with other producers.

Besides, it should be noted that it is possible to have different structures in specific projects at the same stage dependent on their own conditions and characters. For example, some participants may integrate upstream or downstream supply chain to a large extent to ensure a stable cash flow when risks are assessed to be high. Some projects may be operated in a disintegrated model to increase flexibility. Nevertheless, market structure in the same phase represents the characters of the supply chains shaped by most projects.

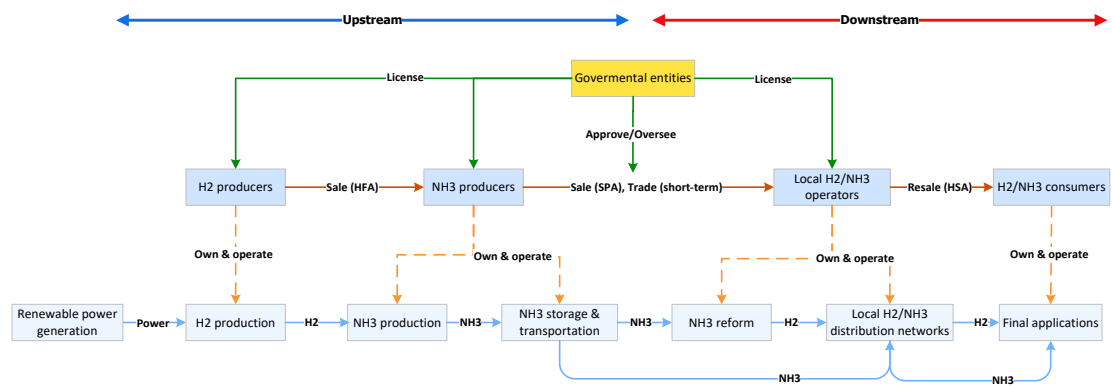

**Fig. 7.** A potential mixed market structure

### 3.2.5 Market structure evaluation

We evaluate the market structures designed for different phases by examining the causes of vertical integration with a case in the global LNG trades. The reasons for the choice are: first, it is not possible to evaluate the market structures by using available data from green ammonia industry, since the industry is not existed yet; second, green ammonia supply chains are highly comparable to that of LNG, especially for the key segments of inter-conversion between gaseous to liquid energy (i.e. natural gas liquefaction and regasification vs. ammonia synthesis and reform); third, according to TCE, transaction costs result in the level of vertical integration, the causes of which should not be specific, but share commonness. The case selected concerns the growth of spot-LNG sales from 1992-2003, due to the support from rich data available, and the fact that spot sales began to develop in global LNG trades since the early 1990s [53]. According to TCE, long-term contracting can be classified as a hybrid governance structure between hierarchy and market structures. Therefore, the growth of spot trades is assumed as a decrease in the level of vertical integration between the upstream and downstream supply chain.

The level of vertical integration is analysed by conducting a multi-linear regression. Table 2 summarizes variables used in the regression model, where $V$ is the dependent variable denoting the level of vertical integration; $C_1$, $C_2$ and $C_3$ are independent variables denoting asset specificity, uncertainty and transaction frequency, respectively. The share of spot trades is used to indicate the level of vertical integration ($V$). Asset specificity ($C_1$) is measured by the share of surplus capacity relative to contract-based trades in the overall liquefaction capacity. It includes spare capacity and capacity employed for spot trades. The average natural gas import price is applied to indicate the degree of uncertainty ($C_2$). The price mostly reflects the price under long-term contracts, since spot trades only accounted for 3.6% of the overall trades on average during this period [53, 54]. In other words, it should have a negative correlation with uncertainty for spot trades. The average price applied is on the basis of import prices in North America and Europe which have higher relativities to the growth of spot trades, due to the industry restructuring and introduction of market pricing mechanism since 1980s and 1990s, respectively [53]. In addition, since the US took up all the trading volume in North America during this period [54], import prices in the US are used to indicate prices in North America. Transaction frequency ($C_3$) is measured by the indicator of annual trading volume. Table 3 summarizes the parameters used in the regression model. The annual data is first standardized with Eq.(1) to eliminate the impacts of data attributes (e.g. dimensions, order of significance, etc.) where $X_n'$ is the standardized data of year $n$ derived from $X_n$. This yields the regression model formulated with standardized sample data shown in Eq.(2), where $β_0$ is the constant; $β_1$, $β_2$, $β_3$ are coefficients; $V_n'$, $C_{1,n}'$, $C_{2,n}'$ and $C_{3,n}'$ are the standardized variables derived from V, $C_1$, $C_1$, and $C_3$, respectively; $e_n$ denotes the residual of year $n$;

Results of the regression are shown in Table 4, 5, and 6. Table 4 summarizes results of joint significance test which reject the null hypothesis at 5% significance level. The model fits the data well ($R^2$ = 95.81%), and the fitted values are compared with actual values in Fig. 8. Table 5 summarizes results of individual significance and multi-collinearlity tests. All the independent variables have passed individual tests at 5% significance level. In addition, all the variance inflation factors (VIFs) are within the tolerance range (VIF < 5), indicating that there is no obvious multi-collinearity between variables. In addition, the regression model is validated by residuals analysis, as shown in Table 6. All the std. residuals are homogenously converged within the range of [-2, 2]. The model has also passed the Anderson-Darling (A-D) test at 5% level of significance. These indicate that residuals are generally independent and subject to the normal distribution.

The results indicate that asset specificity, uncertainty and transaction frequency are associated with the level of vertical integration. Specifically, calls for market support will be intensified with asset specificity and uncertainty reduced, and transaction frequency increased when the market grows. This will finally lead to a transition to a more disintegrated market structure. Conversely, when the industry size is small especially in its infancy, enterprises will most likely to choose hybrid or even hierarchy governance structure to deal with high asset specificity and uncertainty, and low transaction frequency. This will lead to a more integrated market structure. Besides, it should be noted that the vertical integration or disintegration can emerge throughout the supply chain (e.g. a disintegration between production and transportation), not limited to the one mentioned in this case.

$$X_n' = (X_n - \min(X))/(\max(X) - \min(X)), \quad \forall n \in \{1, 2, \dots, N\} \tag{1}$$

$$V_n' = β_0 + β_1 C_{1,n}' + β_2 C_{2,n}' + β_3 C_{3,n}' + e_n, \quad \forall n \in \{1, 2, \dots, N\} \tag{2}$$

**Table 2**

Definition of variables in the multi-regression model

| Notation | Definition | Indicator applied | Unit |
|---|---|---|---|
| $V$ | Vertical integration | Share of spot trades | % |
| $C_1$ | Asset specificity | Share of surplus liquefaction capacity | % |
| $C_2$ | Uncertainty | Average natural gas import price | USD/MBtu |
| $C_3$ | Transaction frequency | Annual trading volume | bcm/yr |

**Table 3**

Parameters used in the multi-regression model

| Parameter (1992-2003) | Value(min) | Value(max) | Unit | Source |
|---|---|---|---|---|
| Share of spot- LNG trades | 1.40 | 7.70 | % | [53, 54] |
| Share of surplus liquefaction capacity | 11.59 | 21.62 | % | [53, 54] |
| Overall LNG trading volume | 80.90 | 149.51 | bcm/yr | [54] |
| Volume of LNG trades in Europe | 20.15 | 38.89 | bcm/yr | [54] |
| Volume of LNG trades in North America | 1.22 | 6.49 | bcm/yr | [54] |
| Natural gas import price in Europe | 1.80 | 4.15 | USD/MBtu | [55] |
| Natural gas import price in the US | 1.50 | 4.40 | USD/MBtu | [56] |

**Table 4**

Results of joint significance test and residuals analysis (5% significance level)

| F-test value | $F_{0.05}(3,7)$ | P | $R^2$ |
|---|---|---|---|
| 53.30 | 4.35 | 0.00 | 95.81% |

**Table 5**

Results of individual significance test (5% significance level) and multi-collinearity test

| Variable | Coefficient | Coefficient value | T-test value | $T_{0.025}(7)$ | VIF value | Tolerated VIF |
|---|---|---|---|---|---|---|
|  | $β_0$ | -0.52 | n.a | 2.36 | n.a | 5 |
| $C_1′$ | $β_1$ | 2.28 | 3.53 |  | 1.18 |  |
| $C_2′$ | $β_2$ | 3.89 | 4.58 |  | 1.95 |  |
| $C_3′$ | $β_3$ | 3.45 | 4.97 |  | 1.94 |  |

**Table 6**

Results of residuals analysis (5% significance level)

| Interval of std. residuals | A-D test value | $A\text{-}D_{0.05}(11)$ |
|---|---|---|
| [-2, 2] | 0.27 | 0.68 |

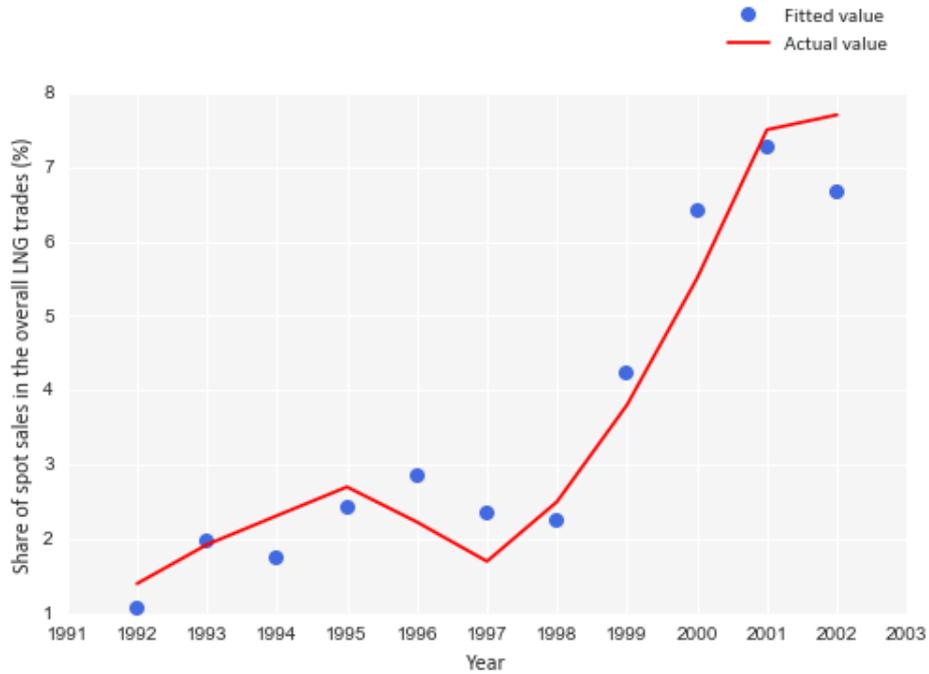

**Fig. 8.** Fitted values compared to actual values

## 3.3 Market emergence

The demand growth is key to the emergence of green ammonia markets. First, targets for the uptake of green ammonia should be set up and involved in the mid or long-term development plans. Similar practices can be observed in some countries or regions. For example, Japan is targeting to expand ammonia fuel use to 4 Mt/yr by 2030 and 30 Mt/yr by 2050 [57]. The ambition of EU is to produce 10 Mt and import 10 Mt of green hydrogen per year by 2030 [58]. The uptake targets can be further subdivided and allocated to multiple sectors, such as power, transport and conventional ammonia sectors.

Second, given that green ammonia is still expensive, government incentives are indispensible to promote the uptake of green ammonia and related technologies. The establishment of incentives can refer to similar practices in other countries and regions. For example, the hydrogen production tax credit issued by the US in 2022 provides 0.6 USD/kg hydrogen produced with a tax credit for 10 years for clean hydrogen producers [59]. The green deal industrial plan recently launched by the EU will carry out a series of auctions for subsidizing green hydrogen production through the European hydrogen bank [60]. Hydrogen producers who win the auctions will receive a fixed premium for each kilogram of hydrogen produced over 10 years [60]. Besides, practices in existing energy industries can also be followed, such as the policy mechanism of feed-in tariff (FIT) to accelerate investment in renewable power generation, the allocation of consumption subsidies for natural gas, etc. [61, 62]. For example, this can be done by applying cost-based pricing in long-term SPAs to incentivise domestic green ammonia production, meanwhile, subsidizing final consumers in the form of tax exemption or premium for the consumption (both domestically produced and imported ammonia). The subsidy for new projects can decrease with production cost reduced over time. In addition, similar to the practice in the EU [58], key technologies and infrastructure deployment can also be funded by either central or local governments to facilitate the supply chain development. In this case, investors can also receive additional benefits.

# 4. Conclusions and policy implications

As a prospective energy carrier and clean fuel, green ammonia is expected to play a key role in the transition to a hydrogen economy. Green ammonia supply chains and associated development were discussed in this paper with a focus on market structure which is key to the supply chain development. The analysis was carried out from a socio-technical perspective, given that supply chains are complex socio-technical systems. Technical system, actors and institutions as well as interactions among them were considered.

(1) The supply chain architecture and associated infrastructure were explored in the light of oil and gas infrastructure and recent studies on hydrogen supply chains. The supply chain comprises of upstream and downstream. The upstream refers to the supply side, from hydrogen production to ammonia transportation and storage, and the downstream concerns the demand side, from ammonia reform (if needed) to end use of hydrogen and ammonia.

(2) Institutional measures to accelerate the market emergence were discussed by referring to practices in some countries and regions towards a hydrogen economy and lessons from the power sector and gas industry. The demand growth is key to the market creation. Hence, uptake targets was proposed to formulated in especially mid and long-term development plans. In addition, certain economic incentives was proposed to introduce to match with the uptake targets. The related practices proposed include tax credit or certain premium to incentivise the supply side, and the mechanism similar to FIT to subsidize the demand side.

(3) Three market structures and related contracts in different development phases were explored based on TCE and lessons from oil and gas industry. An integrated structure was proposed for the infancy. A joint investment or long-term contracts between key segments in both upstream and downstream supply chains, and the creation of bilateral markets between buyers and sellers were proposed to share risks and ensure capital recovery. A disintegrated structure was assumed to be appropriate for the next stage to improve the efficiency. The main practices to restructure the supply chain include providing access to third-parties, and cultivating short-term and spot markets. In the late stage, a competitive structure was proposed towards a more market-driven industry. The main restructuring would be allowing a separation between asset ownership and production activities, and further development of short-term and spot markets. In addition, as mentioned, the designed structures are basic prototypes. Specific mixed structures derived from the three structures were interpreted and proposed to cope with actual conditions.

(4) A multi-regression model was developed to evaluate the proposed market structures with a case study in the LNG industry. The results strongly supported the view from the TCE perspective that a more integrated structure would fit the infancy featuring high asset specificity, high uncertainty and low transaction frequency. However, it would transit to a more disintegrated structure with asset specificity and uncertainty reduced, and transaction frequency increased.

In addition, policy implications resulted from the study are discussed as follows.

(1) We assume the findings and results share commonness on a global scale, and are informative to planning the future green ammonia supply chains and associated markets, but will differ with country in details. In addition, we propose to develop ammonia supply chains to encourage hydrogen energy transition, due to the desirable features of ammonia as an economic and safe energy carrier and clean fuel. In addition, a mature ammonia infrastructure already exist for many decades and can be re-used.

(2) As revealed from the multi-regression process, transaction costs impacted by transaction attributes lead to the level of vertical integration. Therefore, the main findings also apply to market development in other energy sectors. Particularly, the designed market structures and contracts can be used as reference for developing the hydrogen economy. However, the details and differences require further explorations.

Besides, this study has some limitations. It mainly explored market structures and related transitions, however, other aspects and details regarding market and supply chain development need to be further studied. For example, standards and regulations for ammonia production, transportation, reform, etc., detailed institutions for market development, etc. In addition, three regression factors were used in the multi-regression model to evaluate the impact on the level of vertical integration. However, transaction attributes can also be measured by other or more indicators to further improve the estimation.

# Acknowledgements

The author would like to thank China Scholarship Council for the funding provided to carry out this research.

# Nomenclature

| Abbreviation | Full name |
|---|---|
| ALA | Ammonia Logistics Agreement |
| ARA | Ammonia Reform Agreement |
| ASA | Ammonia Synthesis Agreement |
| ATA | Ammonia Transportation Agreement |
| EA | Electrolysis Agreement |
| HDA | Hydrogen Distribution Agreement |
| HFA | Hydrogen Feedstock Agreement |
| HSA | Hydrogen Sales Agreement |
| JVA | Joint Venture Agreement |
| SOE | State-owned enterprise |
| SPA | Sale and purchase agreement |
| TCE | Transaction cost economics |
| TOP | Take-or-pay |

| Index | Definition |
|---|---|
| $n$ | Year |

| Variable/Parameter | Definition | Unit |
|---|---|---|
| $\beta_0$ | Constant of the multi-regression | ea. |
| $\beta_1$ | Coefficient of the multi-regression | ea. |
| $\beta_2$ | Coefficient of the multi-regression | ea. |
| $\beta_3$ | Coefficient of the multi-regression | ea. |
| $C_1$ | Independent variable denoting asset specificity, indicated by the share of surplus liquefaction capacity of natural gas | % |
| $C_{1,n}'$ | Standardized regression factor derived from $C_1$ | ea. |
| $C_2$ | Independent variable denoting uncertainty, indicated by the average natural gas import price | USD/MBtu |
| $C_{2,n}'$ | Standardized regression factor derived from $C_2$ | ea. |
| $C_3$ | Independent variable denoting transaction frequency, indicated by the annual LNG trading volume | bcm/yr |
| $C_{3,n}'$ | Standardized regression factor derived from $C_3$ | ea. |
| $e_n$ | Residual of the multi-regression | ea. |
| $V$ | Dependent variable denoting vertical disintegration, indicated by the share of short-term LNG trades | % |
| $V_n'$ | Standardized regression factor derived from $V$ | ea. |
| $X_n$ | The nth variable of the vector | n.a |
| $X_n'$ | The nth standardized variable of the vector | n.a |